\documentstyle[twocolumn,aps,prl,epsf,psfig,floats]{revtex}
\input{epsf}
\input epsf.sty
\input psfig.sty
\begin{document}

\title{Mott Domains of Bosons Confined on Optical Lattices}

\author{G.G. Batrouni and V. Rousseau}
\address{Institut Non-Lin\'eaire de Nice, Universit\'e de Nice--Sophia
Antipolis, 1361 route des Lucioles, 06560 Valbonne, France}
\author{R.T.~Scalettar}
\address{Physics Department, University of California, Davis, CA 95616}
\author{M. Rigol and A. Muramatsu}
\address{Institut f\"ur Theoretische Physik III, Universit\"at
Stuttgart, Pfaffenwaldring 57, D-70550 Stuttgart, Germany}
\author{P.J.H. Denteneer} 
\address{Lorentz Institute, University of Leiden, P. O. Box 9506,
2300 RA  Leiden, The Netherlands}
\author{M.~Troyer} 
\address{Theoretische Physik, Eidgen\"ossische Technische Hochschule
Z\"urich, CH-8093 Z\"urich, Switzerland}

\address{\mbox{ }} \address{\parbox{14cm}{\rm \mbox{ }\mbox{ }
In the absence of a confining potential, the boson Hubbard model in
its ground state is known to exhibit a superfluid to Mott insulator
quantum phase transition at {\it commensurate} fillings and strong
on-site repulsion. In this paper, we use quantum Monte Carlo
simulations to study the ground state of the one dimensional bosonic
Hubbard model in a trap. We show that some, but not all, aspects of
the Mott insulating phase persist when a confining potential is
present.  The Mott behavior is present for a continuous range of {\it
incommensurate} fillings, a very different situation from the
unconfined case.  Furthermore the establishment of the Mott phase does
not proceed via a quantum phase transition in the traditional sense.
These observations have important implications for the interpretation
of experimental results for atoms trapped on optical lattices.
Initial results show that, qualitatively, the same results persist in
higher dimensions. }}
\address{\mbox{ }}
\address{\parbox{14cm}{\rm \mbox{ }\mbox{ }
PACS numbers: 05.30.Jp, 67.40.Yv, 74.60.Ge, 75.10.Nr
}} \maketitle

\narrowtext

A considerable amount of work has been done in the last decade to
determine the ground state phase diagram of correlated bosons on a
lattice described by the ``boson-Hubbard''
Hamiltonian\cite{FisherM,OURPRL1,Trivedi1,cha,monien}. On-site
repulsion can produce a Mott insulating phase at commensurate
fillings, with a quantum phase transition to a superfluid as the
density is shifted or the interaction strength weakened.  Longer range
interactions can cause charge density wave, stripe, or even supersolid
order\cite{OURPRL3,vanott}. Extensions to disordered systems have
allowed the detailed study of the interplay of randomness and
interactions in quantum
systems\cite{OURPRL2}.

Recently, the trapping of atoms on optical lattices has given another
experimental realization of these bosonic phases.  However, the
quadratic confining potential present in addition to the regular
``lattice'' potential, leads to a number of fundamentally new, and
open, issues: (i) Does the confining potential preclude the formation
of Mott regions by providing a continuous, unbounded, distribution of
local site energies?  (ii) If an insulating phase still exists, how is
it characterized?  (iii) What are the quantitative values of the trap
curvature and interaction strength that support Mott phases?  These
questions are largely unaddressed in the literature.

In this paper we report the first quantum Monte Carlo simulation of
the one-dimensional boson-Hubbard model in a confining quadratic
potential and provide a quantitative map of the state diagram.  To our
knowledge, the only non-mean-field work on this problem\cite{Jaksch}
is on very small systems (five particles). We find that the trap
changes the physics fundamentally from that found in earlier
simulations\cite{OURPRL1} and subsequent analytic\cite{monien}
studies.  For example, the vanishing of the global compressibility,
discussed at length in~\onlinecite{FisherM,OURPRL1} and reported in
recent mean-field studies of the boson-Hubbard model in the context of
optical lattices \cite{Oosten}, but which ignores the confining
potential, is absent. Other recent
papers \cite{Nature,Science,Jaksch,Oosten,Goral} on bosons in optical
traps have likewise emphasized similarities to the physics in the
absence of a confining potential, and have used values of the
unconfined lattice critical coupling to compare with experimental
data.

We find the inhomogeneous potential resulting from the confining trap
is a crucial feature in discussing Mott regions at incommensurate
fillings \cite{Nature}, and the physics of bosons on optical lattices
generally.  The Mott insulating regions exist above a threshold
interaction strength, even without the commensurate filling required
in the non-confined case.  This is a consequence of the inhomogeneous
distribution of boson densities which allows extended Mott domains
with, for example, one particle per site to coexist with regions of
other density and is a unique feature which distinguishes the behavior
of the confined model. Similarly, the global compressibility is
non-zero for all densities, including commensurate ones.  The
different regimes of the state diagram can be characterized by the
topology of the density and local compressibility profiles.
Therefore, in this context, the (global) density, $\rho$, defined as
the total number of particles divided by the system size, loses its
meaning since the particles are not uniformly distributed and the
``system size'' itself is ill-defined. Adding particles can push
bosons deeper into the confining potential at the edges of the system
thus changing the ``size''.

We review briefly the properties of the ground state of the
$d$-dimensional non-confined boson Hubbard model.
This model has two phases: a Mott insulator at commensurate fillings
and sufficiently strong interactions, and a superfluid elsewhere
\cite{FisherM,OURPRL1}.  The critical behavior is of two types: mean
field for transitions induced by tuning the density, and of the
$(d+1)$ dimensional XY universality class when the interaction
strength is swept at fixed commensurate filling. One of the key new
results of this paper will be that this special status of commensurate
filling is {\it lost} in the case of a confining potential since
commensuration itself well defined only locally..

We study the Hamiltonian
\begin{eqnarray}
H&=&-t\sum_{i}(a_{i}^{\dagger}a_{i+1}+ a_{i+1}^{\dagger}a_{i})
\\ \nonumber
&+&V_{0}\sum_{i} n_{i} (n_{i} - 1)
+ V_{c} \sum_{i} (i-L/2)^2 \,\, n_{i},
\end{eqnarray}
at zero temperature.  Here $t$ measures the boson kinetic energy,
$V_0$ the on-site repulsion, $V_c$ the curvature of the quadratic
confining potential and $L$ the number of sites.  In the presence of
the trap, the value of $L$ should be chosen such that for the given
trap curvature, the bosons do not see the edge of the system and
therefore do not leak out.  Our simulations were done with the
world-line quantum Monte Carlo algorithm in the canonical
ensemble \cite{OURPRL1,Worldline}. The chemical potential $\mu=\partial
E / \partial N$ is obtained by differentiating numerically the energy
with respect to the particle number \cite{OURPRB}. In the presence of a
confining potential, it is important to measure the local density of
bosons, $n_i = \langle a^{\dagger}_i a_i \rangle$ as well as the local
compressibility, $\kappa_i = \partial n_i / \partial \mu_i=
\beta [\langle n_i^2 \rangle - \langle n_i \rangle^2 ]$.

\begin{figure}
\psfig{file=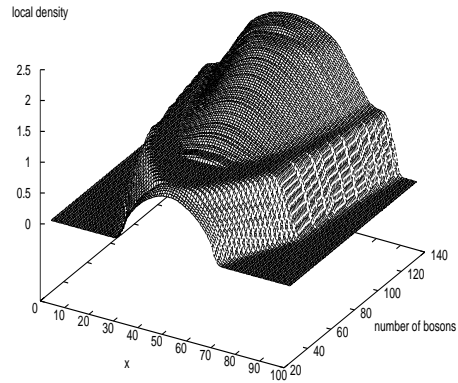,height=7.00cm,width=7cm,angle=-90}
\caption{The evolution of the local density $n_i$ as a
function of increasing total number of bosons.  The trap curvature is
$V_c=0.008$, $L=100$ and the on-site repulsion is $V_0=4$. At low
fillings the system is in a superfluid phase.  Mott insulating
behavior appears as the density is increased, but then at yet larger
fillings a superfluid begins to form at the center of the insulating
region. }
\label{den-prof3d}
\end{figure}

\begin{figure}
\epsfxsize=6.5cm
\epsfysize=5.0cm
\psfig{file=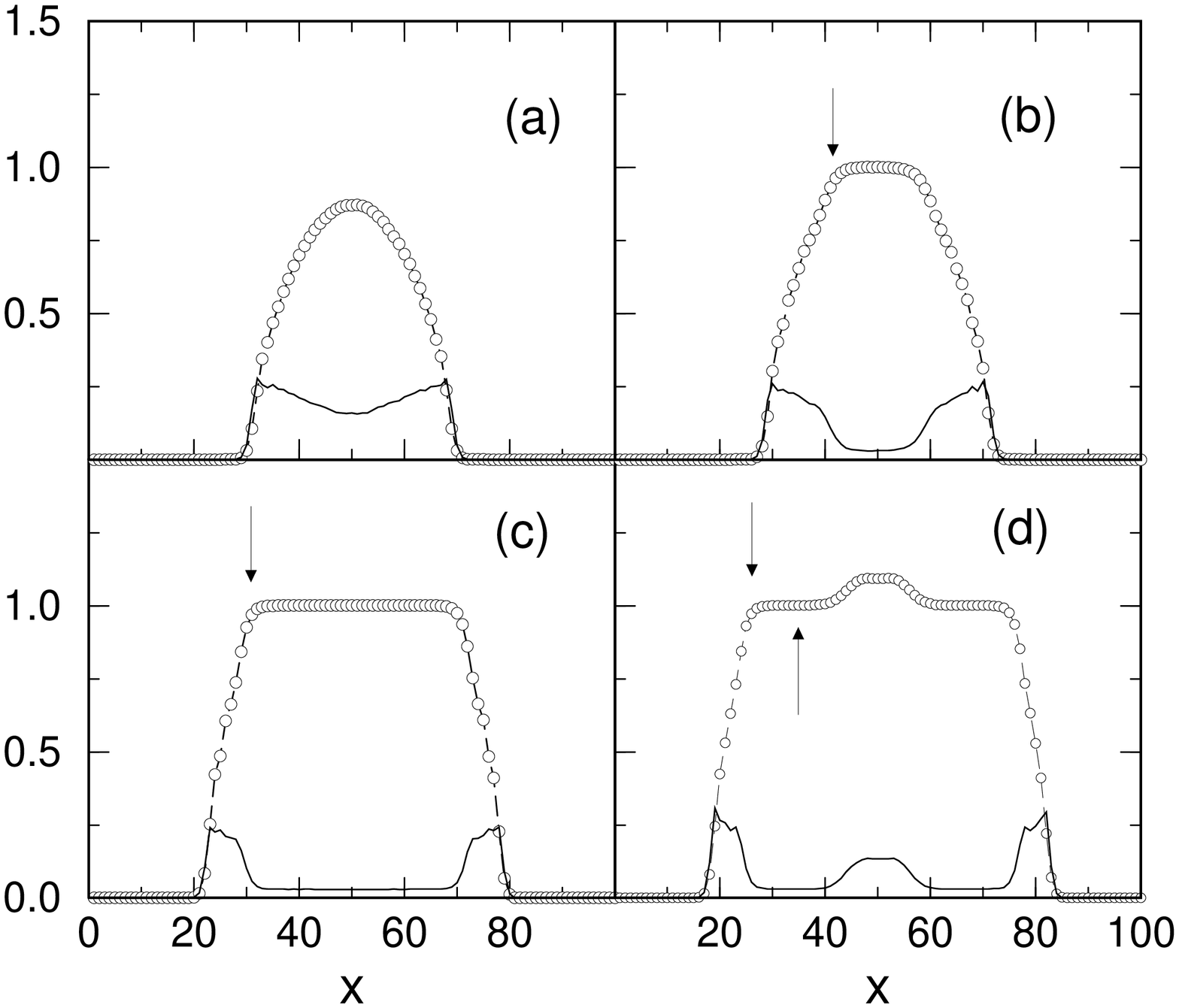,height=5.00cm,width=6.5cm,angle=-90}
\caption{
Cuts across Fig.~1 show the compressibility profile $\kappa_i$ (solid
line) associated with the local density $n_i$ (circles). The fillings
are $N_b=25 {\rm (a)}, 33{\rm (a)}, 50{\rm (a)}$ and $60{\rm
(a)}$. $\kappa_i$ is very small when $n_i=1$. For the arrows see
text.}
\label{slice-prof}
\end{figure}

Figure~\ref{den-prof3d} shows the evolution of the local boson density
with increasing total occupancy of the lattice.  At low fillings the
density profile is smooth, with an inverted parabolic shape reflecting
the confining potential.  Above a critical filling of about 30 bosons
(for this choice of $V_0$ and $V_c$) a plateau with a local filling of
one boson per site develops in the density profile which is analogous
to the Mott structure of $N$ vs $\mu$ in the unconfined model.  This
plateau indicates the presence of an incompressible, insulating region
where $\kappa_i$, as defined above, drops to a small but finite value
(Fig.~\ref{slice-prof}) which vanishes for $V_0\to \infty$. Here, this
behavior of $\kappa_i$ will be taken as the signal for a Mott
region. As the density is increased further, the plateau widens
spatially. But when the energy cost of extending the plateau to
increasingly large values of the confining potential becomes
prohibitive, the occupancy begins to exceed one at the center of the
lattice, indicating a breakdown of Mott behavior there, but {\it not}
everywhere.  Increasing the filling further, for example $N_b=116$,
eventually produces a second Mott region in the center of the system
with a local filling of two bosons per site without destroying totally
the first Mott region. Four slices from Fig.~\ref{den-prof3d} are
shown in Fig.~\ref{slice-prof} along with the local
compressibilities. It is clear that at higher boson numbers, richer
structures where the local compressibility vanishes at several locally
commensurate densities can occur. Mean-field work in two
dimensions\cite{Jaksch} shows a similar coexistence of Mott and
superfluid regions.

\begin{figure}
\epsfxsize=6.5cm
\epsfysize=5.0cm
\psfig{file=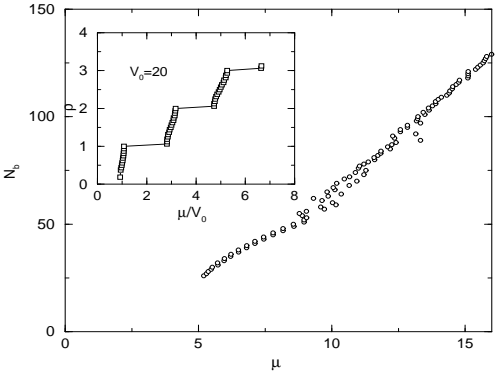,height=5.00cm,width=6.5cm,angle=-90}
\caption{
$N_b$ (number of bosons) as a function of chemical potential, $\mu$
for $V_0=4.5$. No globally incompressible Mott plateau is
observed. Inset shows the unconfined case. }
\label{nboson-mu}
\end{figure}

A central feature of the Mott phase transition of the unconfined
boson-Hubbard model is global incompressibility: A charge gap opens
up, {\it i.e.} the density gets ``stuck'' at $\rho={\rm integer}$ for
a range of chemical potentials $\mu$.  One might, then, crudely
interpret the spatial dependence of the local density in the confined
case, Fig.~\ref{slice-prof}, as rather analogous to the chemical
potential dependence in the unconfined case \cite{Jaksch}. This
assumption, and the $\rho$ versus $\mu$ curve in the unconfined case,
allow us to calculate the site at which a Mott domain is entered or
exited. These are shown as arrows in Fig.~\ref{slice-prof}. However it
is vital to emphasize that while the confined system has locally
incompressible regions, the global compressibility is {\it never} zero
which is seen clearly in Fig.~\ref{nboson-mu}. The main figure should
be contrasted with the nonconfined case (inset).

\begin{figure}
\epsfxsize=6.5cm
\epsfysize=5.0cm
\psfig{file=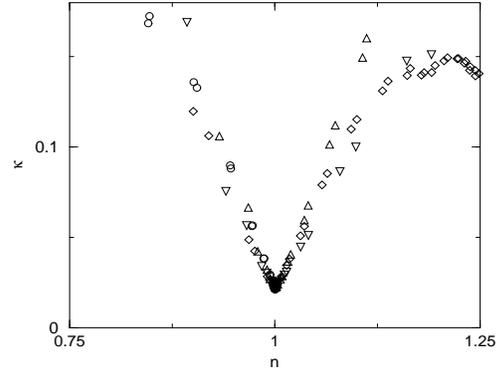,height=5.00cm,width=6.5cm,angle=-90}
\caption{
The local compressibility as a function of local density for $V_0=4$
and $N_b=35$ ($\circ$) and $N_b=80$ ($\diamond$) and for $V_0=4.5$,
$N_b=90$ ($\bigtriangleup$) and $N_b=141$
($\bigtriangledown$). $\kappa$ decreases linearly with $n$ as the Mott
region, $n=1$, is approached from below or above.  This behavior is
dramatically different from the unconfined system where $\kappa$
diverges.}
\label{den-vs-n}
\end{figure}

An important difference in the behavior of the local $\kappa_i$ is
especially evident in one dimension where, in the unconfined case, the
global compressibility diverges\cite{OURPRL1} as the Mott lobe is
approached, $\kappa \propto |\rho - 1|^{-1}$ for the first lobe.
Here, instead, we find $\kappa_i \propto (n_i - 1)$, as shown in
Fig.~\ref{den-vs-n}.  The origin of these differences is of course
that the global compressibility, $\kappa = \beta (\langle \sum_{ij}
n_i n_j \rangle -\langle n \rangle^2)$, probes density correlations at
all length scales.  In the unconfined case, contrary to the confined
system, the establishement of the Mott phase is a true quantum phase
transition: It happens collectively throughout the system and the
correlation length diverges.  There is, however, an interesting
``universality'' in the trapped system as a Mott region is approached.
That is, the values of $\kappa_i$ are the same even as the total
filling and the on-site repulsion are varied (see
Fig.~\ref{den-vs-n}). The same behavior is observed for the $n=2$
locally incompressible phase.

\begin{figure}
\epsfxsize=6.5cm
\epsfysize=5.0cm
\psfig{file=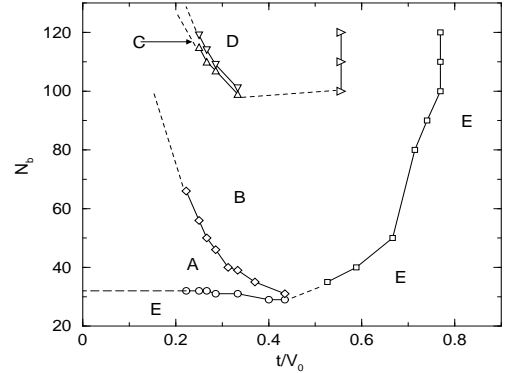,height=5.00cm,width=6.5cm,angle=-90}
\caption{
The state diagram of correlated bosons in a quadratic confining
potential.  The solid lines are to guide the eye, the dashed lines are
extrapolations. See text for details. }
\label{phasediag}
\end{figure}

Sets of runs such as those shown in Fig.~1 allow us to determine the
state diagram as a function of boson filling and interaction strength
for a given trap curvature.  This is shown in Fig.~\ref{phasediag}.
Because of the absence of true phase transitions, we have referred to
Fig.~\ref{phasediag} as a state diagram rather than phase diagram.  As
the filling is increased at fixed interaction strength, one crosses
from a smooth density profile to one which has locally incompressible
``Mott'' domains, if $V_0$ is large enough.  Further increase in the
filling ultimately leads to the formation of regions at the well
center where $n_i > 1$. In Fig.~\ref{phasediag}, region A admits
only\cite{footnote1} locally incompressible regions with $n_i =1$, as
in Fig.~\ref{slice-prof}(b) and (c). Region B has $n_i > 1$ surrounded
by incompressible regions, Fig.~\ref{slice-prof}(d). Region C is where
the central part of the system has an $n_i =2$ incompressible region
which, when the boundary of the system is approached, falls off to a
shoulder of $n_i = 1$ ``Mott'' region before reaching zero
density. The $n_i= 1$ and $n_i = 2$ incompressible regions are
separated by compressible regions. Region D is where the center of the
system is compressible $n_i > 2$, bounded by $n_i = 2$ which in turn
is bounded by $n_i = 1$ incompressible regions. Region E has no
incompressible regions.

Note that the values of $V_0$ at which the A and C regions in
Fig.~\ref{phasediag} are entered are of the same order as those of the
first two true Mott lobes in the nonconfined case \cite{OURPRL1}. This
is consistent with the experimental results on three-dimensional
optical lattices\cite{Nature} which appear to be in agreement with the
expected value in the nonconfined case. Furthermore, the narrowness of
region C could help understand why the experiments\cite{Nature} have
not shown signs of the $n=2$ Mott region, even though $n_i=2.5$ in the
core of the system.

One of the interesting experimental results\cite{Nature} is how
rapidly coherence is re-established when $V_0$ is suddenly reduced
from a value large enough to have produced large incompressible
regions. It was argued that the characteristic time is of the order of
the tunneling (hopping) time between sites \cite{Nature}. This is
entirely consistent with the picture we present here. If, for example,
the density profile is as in Fig.~\ref{slice-prof}(c) when $V_0$ is
suddenly reduced to a very small value, the system will evolve to a
profile like in~\ref{slice-prof}(a) albeit with higher local density
in the center. This is accomplished by particles near the edges
hopping towards the center, and is greatly accelerated by the fact
that the trap is much lower near the center than near edges. In
addition, since there is only one center of nucleation (the geometric
center of the system) there is no slowing down due to competition at
domain walls where different nucleation zones meet.

In summary, we have discussed the nature of locally incompressible
``Mott'' insulating behavior in a one-dimensional system of
interacting bosons in a confining potential at $T=0$.  We conclude
that because of the destruction of translation invariance, great care
should be taken in drawing on the analogy with the unconfined case at
a fundamental level. For one thing, it can support ``Mott'' behavior
off commensurate fillings.  While ``Mott'' regions still exist, the
critical properties are completely altered: Incompressible regions are
established in a very localized way and not at all critically in the
usual sense. These localized regions grow or shrink with $V_0$, but
are always in co-existence with other regions, some compressible and
some which might be incompressible but at higher local integer
filling. In that sense, the formation of these ``Mott'' regions is not
a true quantum critical phenomenon as it is in the unconfined case.

While we have focussed on the new qualitative physics which results
from the confining potential, it is important to emphasize that
experiments on one and two dimensional trapped systems are currently
underway \cite{Science,Greiner,PRIV,cristiani}. For these, our paper
should provide specific quantitative predictions for the critical
ratios of interaction strength to kinetic energy and trap curvature,
as a function of density.  We are currently undertaking these
comparisons.  In addition, in the absence of traps, the phase diagram
is qualitatively the same in one and two
dimensions \cite{FisherM,OURPRL1,Trivedi1,monien}. We expect this to be
true in the confined case too. In fact, initial simulation results in
two dimensions show this to be true.

{\it Note Added:} A QMC study of the three-dimensional system appeared
recently\cite{kash} in which the authors also conclude, like we do,
that one cannot characterize globally the transitions discussed
here. They discuss a signal that can be used experimentally to study
the transition.

We gratefully acknowledge financial support from NSF DMR 9985978, a
CNRS-NSF co-operative grant (France-USA), a PROCOPE grant
(France-Germany), the Stichting FOM and the Swiss National Science 
Foundation. We thank R. Kaiser,
C. Miniatura, T. Pfau and A.B.N. Sue for useful conversations.


\begin{references}
\bibitem{footnote1} Of course, as Fig~\ref{slice-prof} shows, we {\it
always} have compressible regions at the extremities of the
system. Here we are concerned with the interior of the system.
\end{references}
\end{document}